\documentclass[aps,prd,twocolumn,twoside,superscriptaddress,floatfix]{revtex4-1}
\pdfoutput=1

\usepackage{slashed}
\usepackage{amsmath}
\usepackage{multirow}
\usepackage{booktabs}  
\usepackage{graphicx}  
\usepackage{natbib}    
\usepackage{float}

\usepackage{afterpage}
\usepackage{dcolumn}
\usepackage{bm}
\usepackage{amssymb,amsmath}  
\usepackage{lipsum}

\newcommand{\apjl}{ApJL}
\newcommand{\procspie}{Proc.~SPIE}
\newcommand{\physrep}{Phys.~Reps.}
\newcommand{\aap}{A\&A}
\newcommand{\jcap}{JCAP}

\def\bl{\mathbf{l}}

\newcommand{\be}{\begin{equation}}
\newcommand{\ee}{\end{equation}}

\begin{document}
\title{A Small-Scale Modification to the Lensing Kernel}
\author{Boryana Hadzhiyska}
\affiliation{Department of Astrophysical Sciences, Princeton University, Princeton, NJ 08540 USA}
\author{David Spergel}
\affiliation{Department of Astrophysical Sciences, Princeton University, Princeton, NJ 08540 USA}
\affiliation{Flatiron Institute, Simons Foundation, 162 Fifth Avenue, New York, NY 10010, USA}
\author{Joanna Dunkley}
\affiliation{Department of Astrophysical Sciences, Princeton University, Princeton, NJ 08540 USA}
\affiliation{Department of Physics, Princeton University, Princeton, NJ 08544 USA}

\begin{abstract}
Calculations of the Cosmic Microwave Background lensing power implemented into the standard 
cosmological codes such as CAMB and CLASS
usually treat the surface of last scatter as an infinitely thin screen.
However, since the CMB anisotropies
are smoothed out on scales smaller than the diffusion
length due to the effect of Silk damping,
the photons which carry information about the
small-scale density distribution come from slightly earlier times
than the standard recombination time.
The dominant effect is the scale dependence of the 
mean redshift associated with the fluctuations
during recombination. We find that fluctuations at $k = 0.01 {\rm \ Mpc^{-1}}$ 
come from a characteristic redshift of $z \approx 1090$,
while fluctuations at $k = 0.3 {\rm \ Mpc^{-1}}$ 
come from a characteristic redshift of $z \approx 1130$.
We then estimate the corrections to the lensing kernel and
the related power spectra due to this effect.
 We conclude that neglecting it would result in a deviation from the 
true value of the lensing kernel at the half percent level at small CMB scales. 
For an all-sky, noise-free experiment, this 
corresponds to a $\sim 0.1 \sigma$ shift in the observed temperature power
 spectrum on small scales ($ 2500 \lesssim l \lesssim 4000$).
\end{abstract}

\maketitle

\section{Introduction}
\label{chap:intro}
Density fluctuations along the line of sight distort the images of 
observed galaxies. This effect is generally known as gravitational lensing.
 By analyzing such distorted images, one can obtain
a map of the lensing potential, which can then be related to the matter power spectrum
at a given redshift.
In the case of
cosmic microwave background (CMB) photons, these lensing
distortions encode information about the density fluctuations
between the early Universe at $z \approx 1100$
and the present-day Universe. We observe them in the CMB
anisotropies as slight modifications to their statistical properties
\cite{1996ApJ...463....1S,2000PhRvD..62d3007H,PhysRevD.62.063510}.
Over the past decade, cosmologists have measured the CMB lensing signal through 
both auto-correlations and cross-correlations with
other density probes (e.g., cosmic infrared background, galaxy lensing, galaxy counts, 21 cm probes)
\cite{PhysRevD.62.063510,
2007PhRvD..76d3510S,
2012ApJ...753L...9B,
2012PhRvD..86f3519F,
2012PhRvD..86h3006S,
2016A&A...594A..15P}.
Lensing measurements can put constraints on the nature of dark energy 
and the expansion history of the Universe \cite{Hu:1999ek,2009PhRvD..79f5033D,2012JCAP...11..011D}.

Lower-redshift information can be inferred from
weak lensing studies, which measure the distortions of the shapes of
galaxies caused by lensing. In both cases, the goal is to reconstruct
the convergence field, which can be directly related to the projected
matter density by measuring the magnification and shear effects
from either distribution \cite{2006PhR...429....1L}.
Since the lensing reconstruction information is encoded mostly
in the smallest scales observed, high resolution and sensitivity
are crucial for such measurements \cite{2008PhR...462...67M}.

With the improvement in sensitivity
expected in future experiments \cite{doi:10.1117/12.857464,2012SPIE.8452E..1EA,doi:10.1117/12.858314}, 
it is becoming increasingly more important
to take into consideration corrections to 
 the observed power spectra, which have until now been negligible.
 One such effect comes
from the fact that on scales smaller than
the 
diffusion length, the density anisotropies
are smoothed out due to photon diffusion damping. This means that
photons carrying small-scale information
 need to have come from slightly earlier times
than the standard recombination time \cite{1968ApJ...151..459S,1970ApJ...162..815P}.
The dominant effect is the scale dependence of the 
mean redshift associated with the fluctuations
during recombination.

The standard calculation of the CMB lensing power implemented into 
numerical codes such as CAMB \cite{2011ascl.soft02026L} and CLASS \cite{2011JCAP...07..034B} treats
 the surface of last scatter as an infinitely thin screen.
In this paper, we provide a modified
estimation of the distance to last scattering as a function of scale.
This correction takes into
account the scale dependence of the 
recombination redshift in the calculation of the 
lensing kernel, which is needed to obtain the lensing power spectrum.
We finally  evaluate
the percentage difference in the lensing kernel resulting from this modification and discuss its
significance
given the
expected sensitivity of future experiments.

\section{Effect on Lensing Kernel}
\begin{figure}[t]
      \centering
      \includegraphics[width=3.5in]{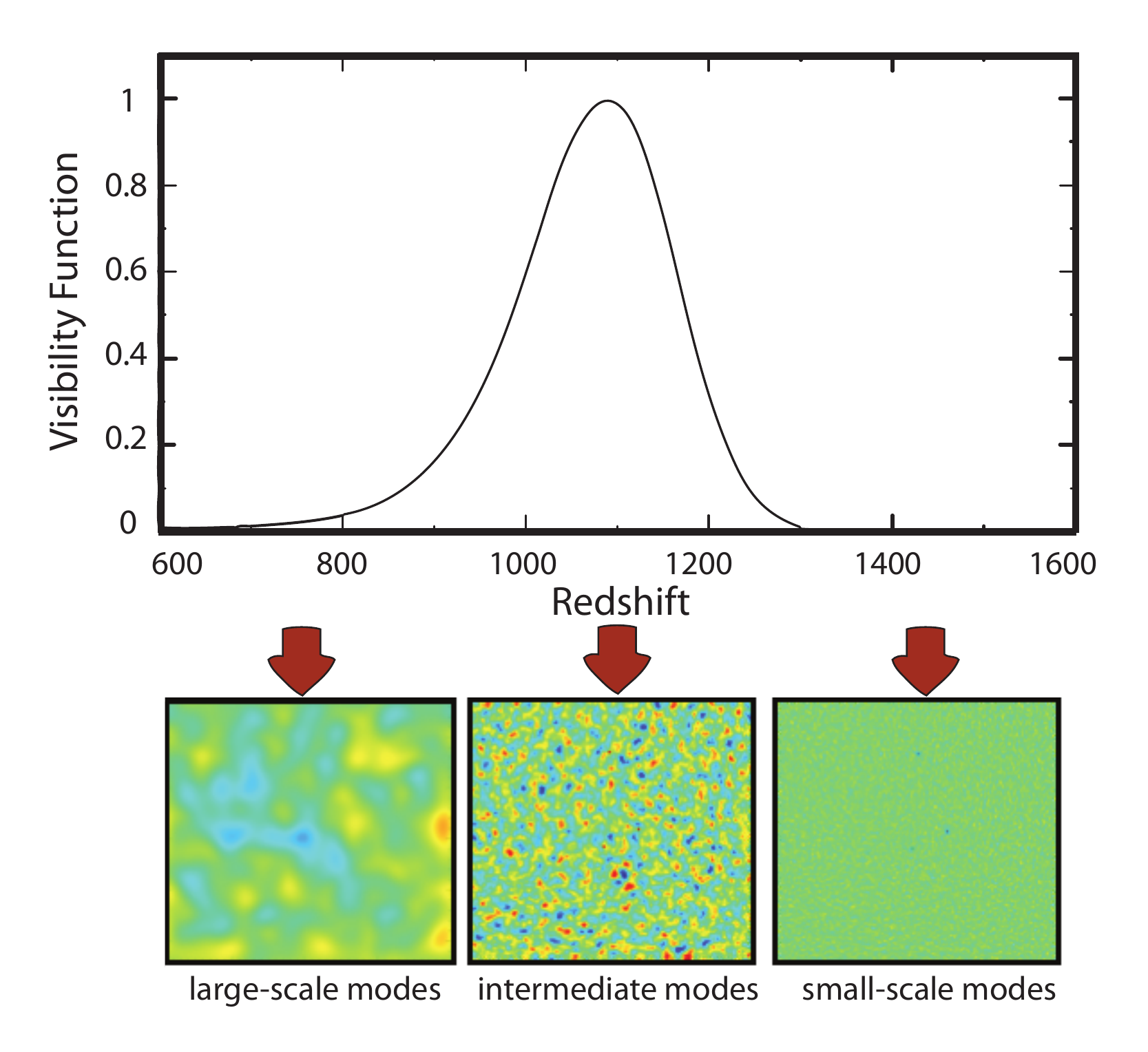}
      \caption{The CMB anisotropies can be broken into small-scale, intermediate-scale
        and large-scale. Due to  diffusion damping, the small-scale information
        is provided by photons which last scattered at redshifts larger than the redshift of 
        recombination. In contrast, at smaller redshifts, the information on small scales is lost,
      and large-scale information is provided by late-time photons.}
      \label{fig:draw}
\end{figure}

\begin{figure}[ht]
      \centering
      \includegraphics[width=3.7in]{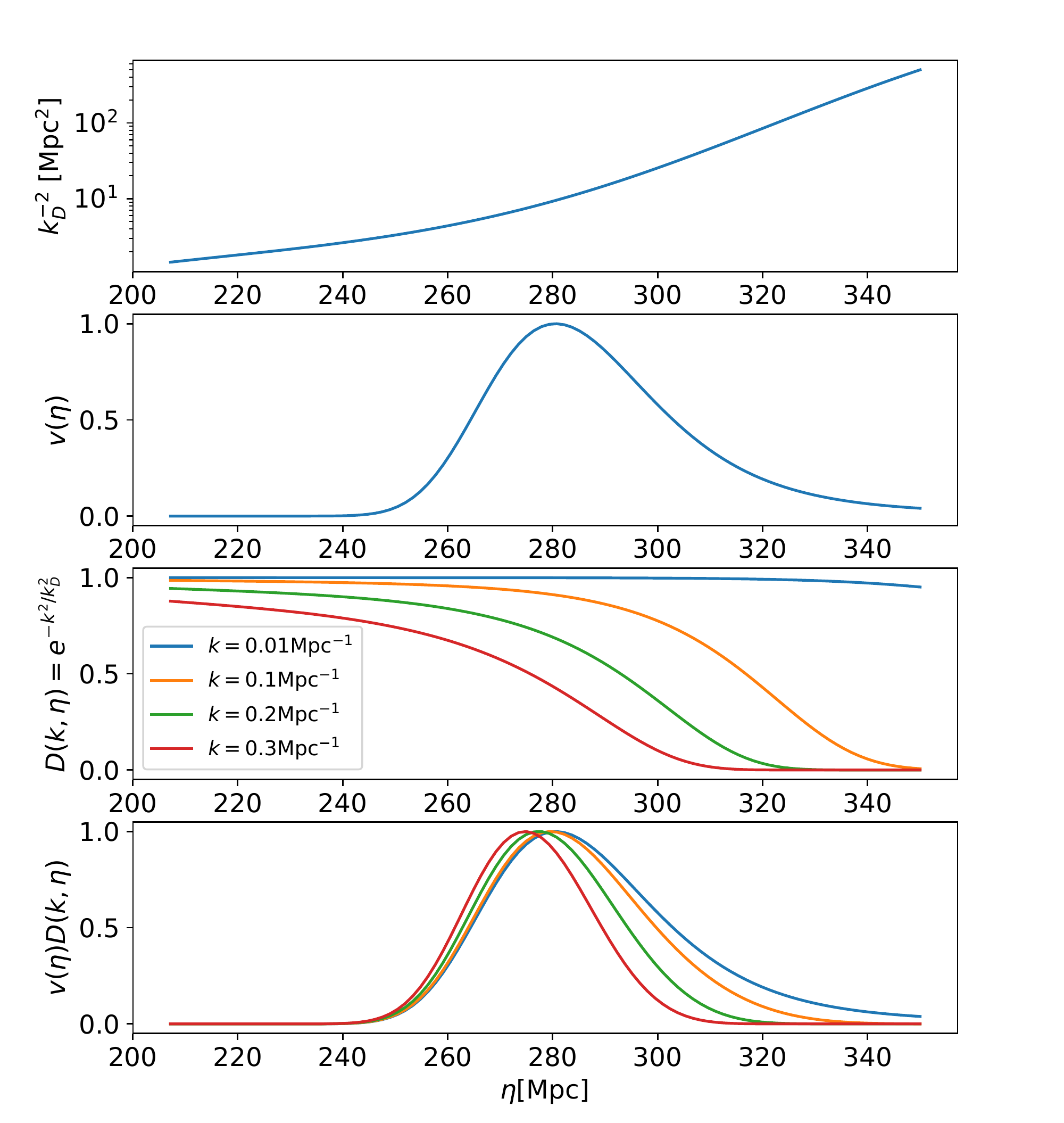}
      \caption{\textbf{Top two panels:} Damping scale and visibility function as a function of conformal time. 
	\textbf{Bottom two panels:}
	Damping factor and the normalized product between the damping factor and the visibility function for
	 different wavenumbers $k$. Smaller-scale modes are most likely to have scattered at earlier times.}
      \label{fig:D_k}
\end{figure}

The weighted projection of the matter density contrast $\delta$, known as the
convergence field, encodes information about the density fluctuations
in the Universe since the period of recombination and is expressed as:
\begin{equation}
  \kappa(\hat n) = \int_0^\infty d z W^\kappa(z) \delta (\chi(z) \hat n, z),
\end{equation}
where $\chi(z)$ is the conformal distance between us and some event at 
redshift $z$ \cite{2001PhR...340..291B}. In a flat universe, the lensing kernel $W^\kappa$ is given by:
\begin{equation}
W^\kappa(z) = \frac{3}{2} \Omega_m H_0^2 \frac{1+z}{H(z)} \ \frac{\chi(z)}{c}
\int_z^\infty d z_s p_s(z_s)\frac{\chi(z_s)-\chi(z)}{\chi(z_s)} ,
\end{equation}
where $p_s(z)$ is the normalized distribution of sources as
a function of redshift.

In the case of the CMB, it is standard to assume that the photons
come predominantly from the redshift of recombination 
$z(\eta_\ast) \equiv z_\ast$, so that one can approximate the source
distribution as $p_s \approx \delta_D(z-z_\ast)$ and thus obtain
the kernel \cite{2006PhR...429....1L}: 
\begin{equation}
W^\kappa(z) = \frac{3}{2} \Omega_m H_0^2 \frac{1+z}{H(z)} \ \frac{\chi(z)}{c}
\frac{\chi(z_\ast)-\chi(z)}{\chi(z_\ast)} .
\end{equation}
We, thus, see that the usual approach for calculating the lensing kernel, also employed by
the cosmological codes CLASS and CAMB,
treats the surface of last scatter as an infinitely thin screen. 
However, the CMB photons come from a range of redshifts which peaks at the period
of recombination. The photons which last scattered at earlier times contain
more small-scale information than those coming from later times
because as the diffusion damping scale increases with time, anisotropies
are smoothed out and information on small scales is lost \cite{1968ApJ...151..459S,1970ApJ...162..815P}.
The main effect is that the 
mean redshift associated with the fluctuations
during recombination becomes scale-dependent.
This effect is illustrated in Fig.
\ref{fig:draw}.

This claim can be supported quantitatively by considering the visibility
function $v(\eta)$, which expresses the most probable time at which a CMB photon last scattered, and
the damping factor $\exp[-k^2/k_D(\eta)^2]$,
which measures how much the growth of a given mode is suppressed as a function
of time \cite{Hu:1995em}. Their product, computed for each mode, informs us about the most likely
time at which the photons encoding information on the given mode last scattered:
\begin{equation}
\label{eq:eta_stk}
g(k,\eta) = e^{-k^2/k_D(\eta)^2} v(\eta) . 
\end{equation}

As seen in the lowest panel in Fig. \ref{fig:D_k}, for small-scale modes, the product 
between the visibility function and the damping factor peaks at earlier
times than the standard recombination time
 $\eta_\ast$, which shows that the CMB photons providing information
 on small-scale anisotropies ($k \gtrsim 0.1 \rm{Mpc^{-1}}$) are
more likely to have come from an earlier time than the mean  recombination time.
On even smaller scales ($k \gtrsim 0.3 \rm{Mpc^{-1}}$), where the primary anisotropies are washed out,
 the effective emission time is shifted to later times, as the motion of the photon-baryon fluid
 starts to be dominated by its infall into the CDM potential wells during matter domination. 
In this regime, the approximation which we are adopting breaks down and more careful analysis
 is needed. However, since we are interested in the effect on the temperature and the polarization 
power spectra for $l \lesssim 4000$, we can neglect the baryon effect.

\begin{figure}[H]
      \centering
      \includegraphics[width=3.5in]{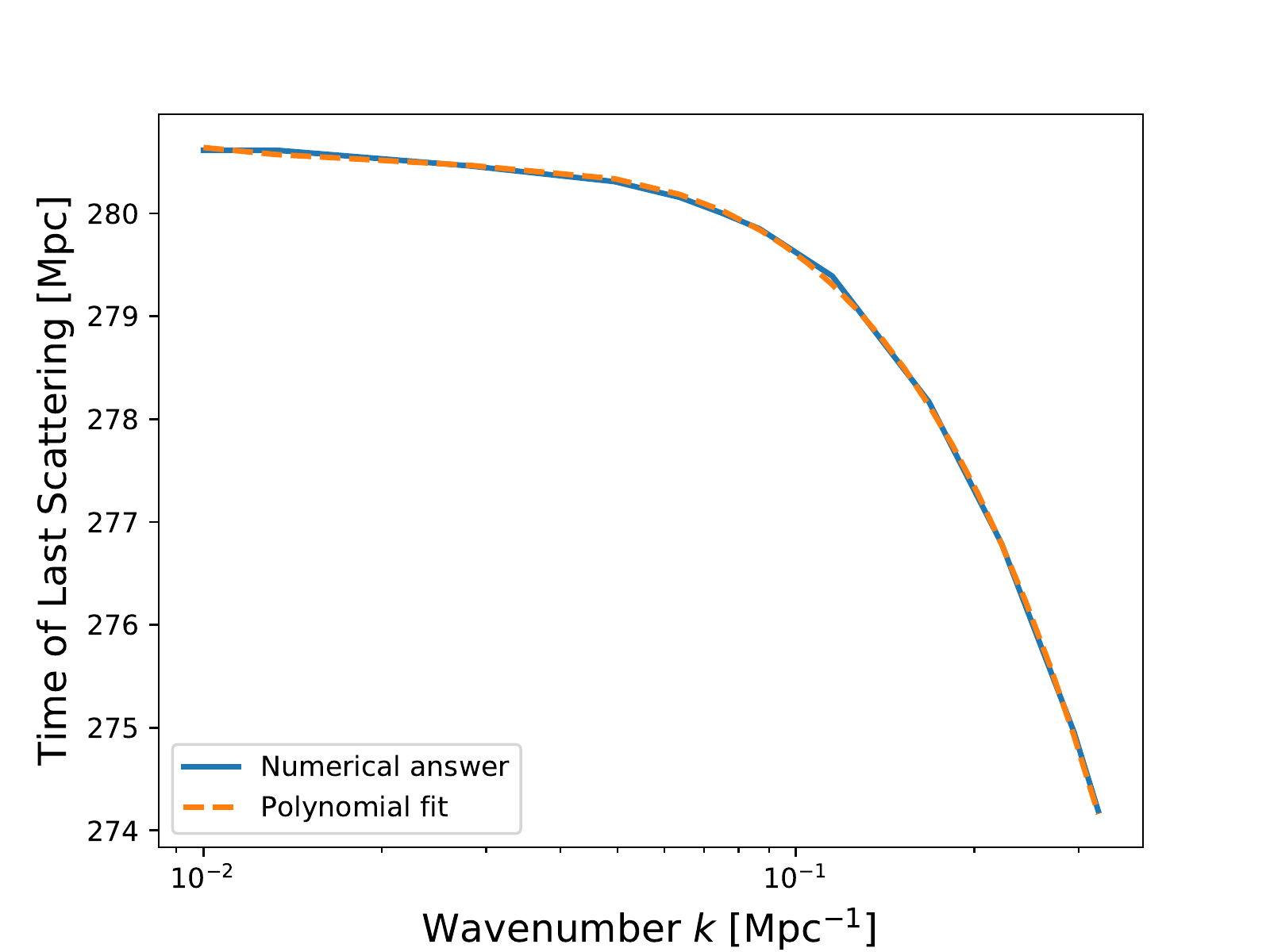}
      \caption{Time of the last scattering of photons as a function of
      CMB scale. The blue curve is derived numerically by computing
      the peak of 
      the product $g(k,\eta)=D(k,\eta)v(\eta)$ for each wavenumber $k$.
      The orange curve uses our approximate result from
      Eq. \ref{eq:eta_stk}.}
      \label{fig:etastk}
\end{figure}

A plot  of conformal time  $\eta_\ast(k)$ $[\rm{Mpc}]$  versus
wavenumber $k$ $[\rm{Mpc^{-1}}]$ obtained
by numerically computing the peak position for each mode is
 shown in Fig. \ref{fig:etastk}. We fit a cubic polynomial to this function,
finding the form:
\begin{eqnarray}
 \eta_\ast(k) = -2.14[\ln(k)]^4 -15.67 [\ln(k)]^3  \nonumber \\ -42.46[\ln(k)]^2 
-50.77 [\ln(k)]+257.76
\end{eqnarray}

We can now incorporate the scale-dependence of $\eta_\ast(k)$ into the kernel and obtain
its form as a function of both redshift and CMB scale $k$, assuming that for each $k$, the source
distribution can be approximated by $p_s \approx \delta_D(\eta - \eta_\ast(k))$:
\begin{equation}
W^\kappa(z,k) = \frac{3}{2} \ \Omega_m H_0^2 \frac{1+z}{H(z)} \ \frac{\chi(z)}{c}
\frac{\chi(z_\ast,k)-\chi(z)}{\chi(z_\ast,k)} .
\end{equation}

\begin{figure}
      \centering
      \includegraphics[width=3.5in]{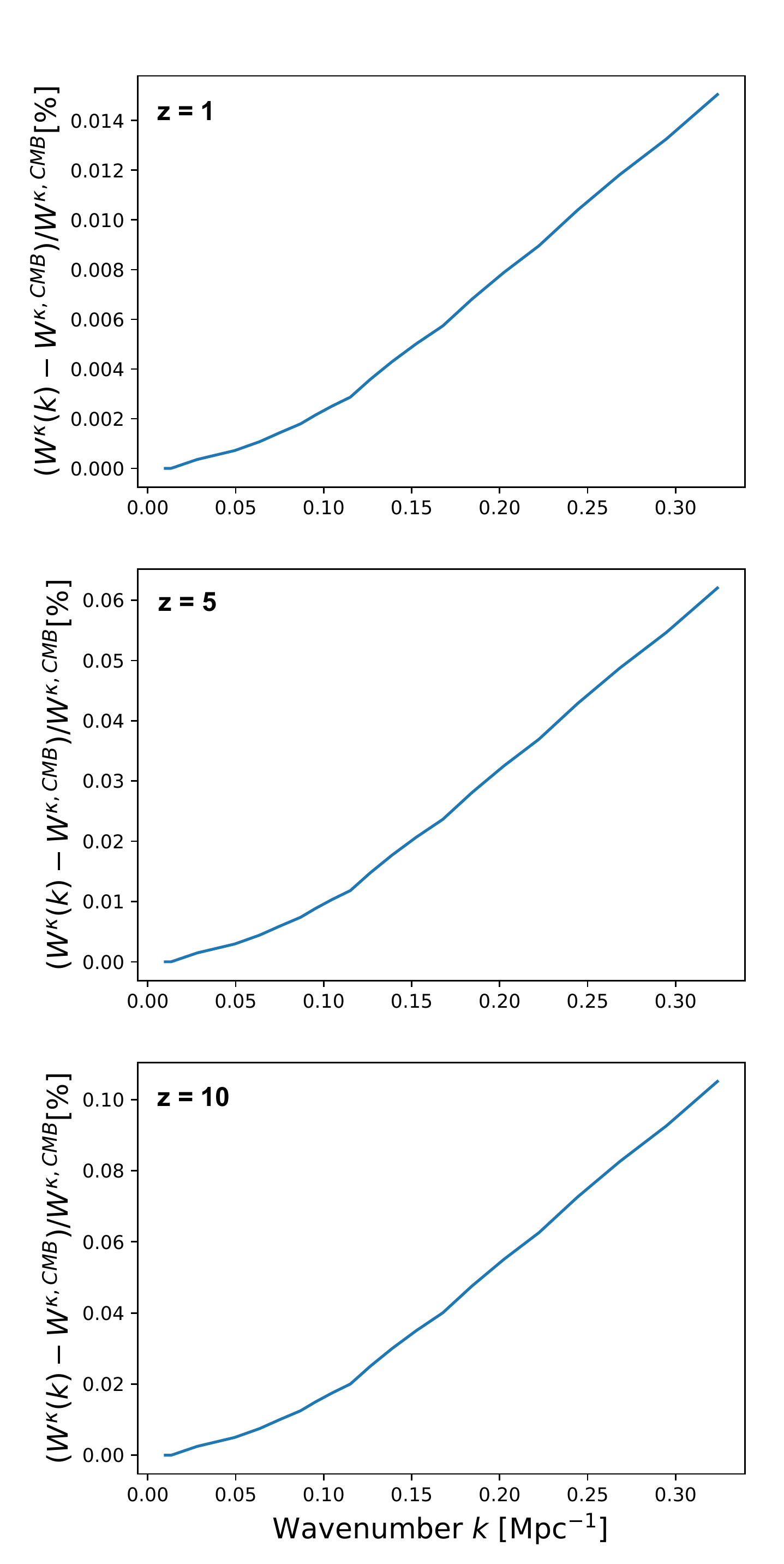} 
     \caption{Fractional difference in the lensing kernel at redshifts $z = 1$, $z = 5$ and $z = 10$ 
        assuming that the distance to last scattering
        depends on the CMB scale $k$. The deviations from the standard
	value of the kernel increase with redshift  and decrease with scale.} 
      \label{fig:lens_z}
\end{figure}

We show the impact on the lensing kernel as a function of CMB scale ($k$) in
three panels (Fig.
\ref{fig:lens_z}),
each of which is a snapshot
at a given redshift: $z = 1$, $z = 5$, and $z = 10$, respectively. One sees that the
fractional differences is increasing approximately linearly with $k$ due to
 the fact that the largest deviations of $\eta_\ast(k)$ from the
standard value arise at the smaller scales ($k \sim 0.3 {\rm \ Mpc^{-1}}$).
Another observation is that the deviation from the standard
value of the kernel increases with redshift:
 at redshift $z = 1$, the fractional difference is merely 0.01\%
on small scales,
but it reaches 0.1\% for redshift $z = 10$. 

Using the Limber approximation \cite{1954ApJ...119..655L,1992ApJ...388..272K},
the modified equation for the power spectrum of the convergence field due to CMB lensing
becomes:
\begin{equation}
C_{l,l'}^{\kappa \kappa} = \int^\infty_0 \frac{d z}{c} \frac{H(z)}{\chi(z)^2} W^\kappa(z,k'=l' / \chi_\ast
)^2 P (k=l / \chi, z),
\end{equation}
where $\chi_\ast$ is the mean distance to the last scattering surface. Note that in contrast with the standard
calculation, here the lensing power spectrum depends on the two scales $l$ ($k$) and $l'$ ($k'$). Thus, 
$C_{l,l'}^{\kappa \kappa}$ effectively describes how much our signal will be lensed for a lens of size
$l$ ($k$) and a given size of the CMB anisotropy $l'$ ($k'$).

Since we expect very high sensitivity
in the future measurements of the temperature power spectrum, an interesting observable to consider is the lensed temperature power spectrum $C_l^{\tilde T \tilde T}$, which is approximately given by the following modified equation:
\begin{eqnarray}
C_l^{\tilde T \tilde T} \approx 
\int \frac{d^2 \bl'}{(2 \pi)^2} \frac{4[\bl' \cdot (\bl-\bl')]^2}{|\bl-\bl'|^4} C_{|\bl-\bl'|,l'}^{\kappa \kappa} C_{l'}^{TT} 
\nonumber \\
+C_l^{TT} \Big[ 1- \frac{1}{\pi} l^2 \int \frac{d l'} {l'} C_{l',l}^{\kappa \kappa} \Big] .
\label{eq:TT}
\end{eqnarray}

\begin{figure}[H] 
      \centering
      \includegraphics[width=3.5in]{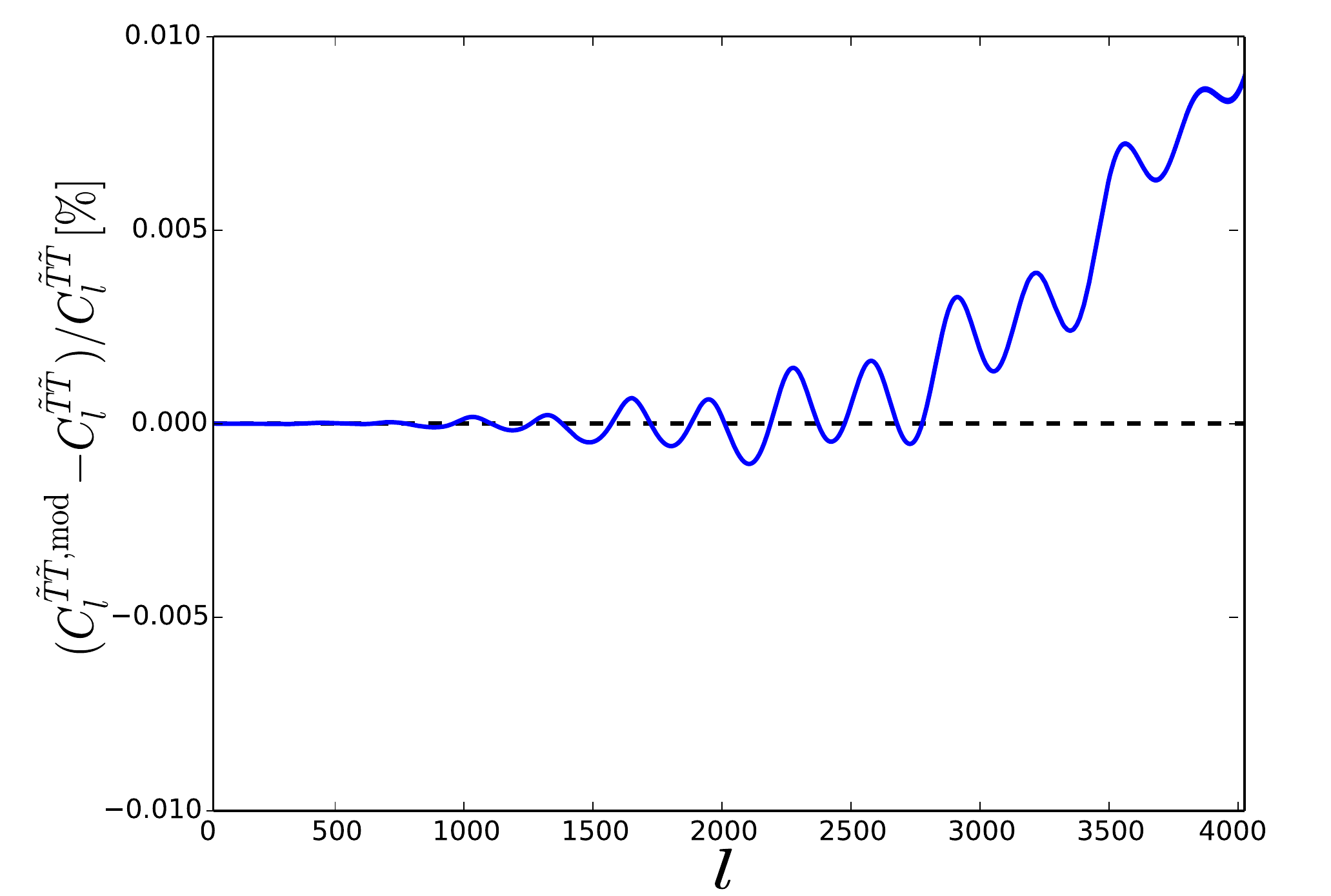}
      \caption{Percentage difference in the lensed temperature power spectrum assuming a 
	scale-dependent distance to last scattering. This result is derived by using the full-sky non-perturbative
	approximation in Ref. \cite{Challinor:2005jy} to second order.}
      \label{fig:TTfrac}
\end{figure}

The resulting fractional difference in the temperature
power spectrum
is shown in Fig. \ref{fig:TTfrac}, where we use a full-sky non-perturbative
 approximation \cite{Challinor:2005jy}.
As we increase $l'$ and thus the distance
to the source (small scales come from earlier times), the overall amplitude of the difference
 also gets larger, as expected in lensing theory. 
The average percentage difference across the small-scale modes $2500 \lesssim l \lesssim 4000$ is 0.004\%.
Consequently, the measured temperature power spectrum 
in an idealized all-sky, noise-free experiment with a signal-to-noise ratio of $\approx 2200$ on these scales 
would be shifted by $\sim 0.1 \sigma$ from the theoretically predicted one.

The corresponding equation for the lensed  B-mode polarization power
spectrum is:
\be
C_l^{\tilde B \tilde B} \approx \int \frac{d^2 \bl'}{(2 \pi)^2} \frac{4[\bl' \cdot (\bl-\bl')]^2}{|\bl-\bl'|^4} \sin^2(2 \phi_{\bl,\bl'}) C_{|\bl-\bl'|,l'}^{\kappa \kappa} C_{l'}^{EE} .
\ee
Since the B-mode power
spectrum gives us a direct probe of the lensing amplitude, the resulting fractional difference
is larger than in the case of temperature, peaking at roughly $l \approx 3000$ with
$\Delta C_l^{\tilde B \tilde B} / C_l^{\tilde B \tilde B} \approx 0.03\%$. 
However, this deviation
would be harder to measure in the near future due to the lower sensitivity of the
polarization measurements
compared with the temperature. 

\section{Conclusion}
\label{chap:conc}
The smaller-scale anisotropies
observed in the CMB come from slightly earlier
times, which implies that the time of photon last 
scattering is dependent  on the physical scale.
This has important implications for the lensing kernel
used to compute the observable power spectra.
We found differences in the lensing kernel of
 $\sim 0.1 \%$ at redshift $z = 10$ and
of $\sim 0.06 \%$ at redshift $z = 5$ for the
smaller scales ($k \sim 0.3 {\rm Mpc^{-1}}$). 
Consequently, this leads to an average deviation of
0.004\% in the temperature power spectrum on
small scales $2500 \lesssim l \lesssim 4000$.
In the future, experiments will ideally only be limited by cosmic
variance and will thus 
measure the temperature power spectrum 
with a signal-to-noise ratio of $\approx 2200$ on scales
 $2500 \lesssim l \lesssim 4000$. 
Neglecting the scale-dependence of recombination then
 would lead  to a measurable deviation from the predicted power spectrum
of about $\sim 0.1 \sigma$ on these scales. An observable for which we expect the effect to be
more prominent in the near future is the reconstructed lensing
power spectrum, as it depends on the 4-point function of the lensed temperature map.
We are hoping to look into it in a future paper.
In the current age of precision cosmology,
implementing such
sub-percent modifications to the 
observable power spectra is becoming
increasingly important.

\begin{acknowledgments}
We would like to thank Mathew S. Madhavacheril for providing us with
 useful lensing computational tools. We are further grateful to  Antony Lewis, Anthony Challinor and Blake Sherwin
for pointing out and discussing an inaccuracy in our application of the effect to CMB lensing.
 B.~H. would like to thank the Department of
Astrophysical Sciences at Princeton for the immense support during her undergraduate degree.
\end{acknowledgments}

\bibliographystyle{apsrev4-1} 
\bibliography{kernel} 

\begin{thebibliography}{25}%
\makeatletter
\providecommand \@ifxundefined [1]{%
 \@ifx{#1\undefined}
}%
\providecommand \@ifnum [1]{%
 \ifnum #1\expandafter \@firstoftwo
 \else \expandafter \@secondoftwo
 \fi
}%
\providecommand \@ifx [1]{%
 \ifx #1\expandafter \@firstoftwo
 \else \expandafter \@secondoftwo
 \fi
}%
\providecommand \natexlab [1]{#1}%
\providecommand \enquote  [1]{``#1''}%
\providecommand \bibnamefont  [1]{#1}%
\providecommand \bibfnamefont [1]{#1}%
\providecommand \citenamefont [1]{#1}%
\providecommand \href@noop [0]{\@secondoftwo}%
\providecommand \href [0]{\begingroup \@sanitize@url \@href}%
\providecommand \@href[1]{\@@startlink{#1}\@@href}%
\providecommand \@@href[1]{\endgroup#1\@@endlink}%
\providecommand \@sanitize@url [0]{\catcode `\\12\catcode `\$12\catcode
  `\&12\catcode `\#12\catcode `\^12\catcode `\_12\catcode `\%12\relax}%
\providecommand \@@startlink[1]{}%
\providecommand \@@endlink[0]{}%
\providecommand \url  [0]{\begingroup\@sanitize@url \@url }%
\providecommand \@url [1]{\endgroup\@href {#1}{\urlprefix }}%
\providecommand \urlprefix  [0]{URL }%
\providecommand \Eprint [0]{\href }%
\providecommand \doibase [0]{http://dx.doi.org/}%
\providecommand \selectlanguage [0]{\@gobble}%
\providecommand \bibinfo  [0]{\@secondoftwo}%
\providecommand \bibfield  [0]{\@secondoftwo}%
\providecommand \translation [1]{[#1]}%
\providecommand \BibitemOpen [0]{}%
\providecommand \bibitemStop [0]{}%
\providecommand \bibitemNoStop [0]{.\EOS\space}%
\providecommand \EOS [0]{\spacefactor3000\relax}%
\providecommand \BibitemShut  [1]{\csname bibitem#1\endcsname}%
\let\auto@bib@innerbib\@empty
\bibitem [{\citenamefont {{Seljak}}(1996)}]{1996ApJ...463....1S}%
  \BibitemOpen
  \bibfield  {author} {\bibinfo {author} {\bibfnamefont {U.}~\bibnamefont
  {{Seljak}}},\ }\href {\doibase 10.1086/177218} {\bibfield  {journal}
  {\bibinfo  {journal} {\apj}\ }\textbf {\bibinfo {volume} {463}},\ \bibinfo
  {pages} {1} (\bibinfo {year} {1996})},\ \Eprint
  {http://arxiv.org/abs/astro-ph/9505109} {astro-ph/9505109} \BibitemShut
  {NoStop}%
\bibitem [{\citenamefont {{Hu}}(2000)}]{2000PhRvD..62d3007H}%
  \BibitemOpen
  \bibfield  {author} {\bibinfo {author} {\bibfnamefont {W.}~\bibnamefont
  {{Hu}}},\ }\href {\doibase 10.1103/PhysRevD.62.043007} {\bibfield  {journal}
  {\bibinfo  {journal} {\prd}\ }\textbf {\bibinfo {volume} {62}},\ \bibinfo
  {eid} {043007} (\bibinfo {year} {2000})},\ \Eprint
  {http://arxiv.org/abs/astro-ph/0001303} {astro-ph/0001303} \BibitemShut
  {NoStop}%
\bibitem [{\citenamefont {Zaldarriaga}(2000)}]{PhysRevD.62.063510}%
  \BibitemOpen
  \bibfield  {author} {\bibinfo {author} {\bibfnamefont {M.}~\bibnamefont
  {Zaldarriaga}},\ }\href {\doibase 10.1103/PhysRevD.62.063510} {\bibfield
  {journal} {\bibinfo  {journal} {Phys. Rev. D}\ }\textbf {\bibinfo {volume}
  {62}},\ \bibinfo {pages} {063510} (\bibinfo {year} {2000})}\BibitemShut
  {NoStop}%
\bibitem [{\citenamefont {{Smith}}\ \emph {et~al.}(2007)\citenamefont
  {{Smith}}, \citenamefont {{Zahn}},\ and\ \citenamefont
  {{Dor{\'e}}}}]{2007PhRvD..76d3510S}%
  \BibitemOpen
  \bibfield  {author} {\bibinfo {author} {\bibfnamefont {K.~M.}\ \bibnamefont
  {{Smith}}}, \bibinfo {author} {\bibfnamefont {O.}~\bibnamefont {{Zahn}}}, \
  and\ \bibinfo {author} {\bibfnamefont {O.}~\bibnamefont {{Dor{\'e}}}},\
  }\href {\doibase 10.1103/PhysRevD.76.043510} {\bibfield  {journal} {\bibinfo
  {journal} {\prd}\ }\textbf {\bibinfo {volume} {76}},\ \bibinfo {eid} {043510}
  (\bibinfo {year} {2007})},\ \Eprint {http://arxiv.org/abs/0705.3980}
  {arXiv:0705.3980} \BibitemShut {NoStop}%
\bibitem [{\citenamefont {{Bleem}}\ \emph {et~al.}(2012)\citenamefont
  {{Bleem}}, \citenamefont {{van Engelen}}, \citenamefont {{Holder}},
  \citenamefont {{Aird}}, \citenamefont {{Armstrong}}, \citenamefont {{Ashby}},
  \citenamefont {{Becker}}, \citenamefont {{Benson}}, \citenamefont
  {{Biesiadzinski}}, \citenamefont {{Brodwin}}, \citenamefont {{Busha}},
  \citenamefont {{Carlstrom}}, \citenamefont {{Chang}}, \citenamefont {{Cho}},
  \citenamefont {{Crawford}}, \citenamefont {{Crites}}, \citenamefont {{de
  Haan}}, \citenamefont {{Desai}}, \citenamefont {{Dobbs}}, \citenamefont
  {{Dor{\'e}}}, \citenamefont {{Dudley}}, \citenamefont {{Geach}},
  \citenamefont {{George}}, \citenamefont {{Gladders}}, \citenamefont
  {{Gonzalez}}, \citenamefont {{Halverson}}, \citenamefont {{Harrington}},
  \citenamefont {{High}}, \citenamefont {{Holden}}, \citenamefont
  {{Holzapfel}}, \citenamefont {{Hoover}}, \citenamefont {{Hrubes}},
  \citenamefont {{Joy}}, \citenamefont {{Keisler}}, \citenamefont {{Knox}},
  \citenamefont {{Lee}}, \citenamefont {{Leitch}}, \citenamefont {{Lueker}},
  \citenamefont {{Luong-Van}}, \citenamefont {{Marrone}}, \citenamefont
  {{Martinez-Manso}}, \citenamefont {{McMahon}}, \citenamefont {{Mehl}},
  \citenamefont {{Meyer}}, \citenamefont {{Mohr}}, \citenamefont {{Montroy}},
  \citenamefont {{Natoli}}, \citenamefont {{Padin}}, \citenamefont {{Plagge}},
  \citenamefont {{Pryke}}, \citenamefont {{Reichardt}}, \citenamefont {{Rest}},
  \citenamefont {{Ruhl}}, \citenamefont {{Saliwanchik}}, \citenamefont
  {{Sayre}}, \citenamefont {{Schaffer}}, \citenamefont {{Shaw}}, \citenamefont
  {{Shirokoff}}, \citenamefont {{Spieler}}, \citenamefont {{Stalder}},
  \citenamefont {{Stanford}}, \citenamefont {{Staniszewski}}, \citenamefont
  {{Stark}}, \citenamefont {{Stern}}, \citenamefont {{Story}}, \citenamefont
  {{Vallinotto}}, \citenamefont {{Vanderlinde}}, \citenamefont {{Vieira}},
  \citenamefont {{Wechsler}}, \citenamefont {{Williamson}},\ and\ \citenamefont
  {{Zahn}}}]{2012ApJ...753L...9B}%
  \BibitemOpen
  \bibfield  {author} {\bibinfo {author} {\bibfnamefont {L.~E.}\ \bibnamefont
  {{Bleem}}}, \bibinfo {author} {\bibfnamefont {A.}~\bibnamefont {{van
  Engelen}}}, \bibinfo {author} {\bibfnamefont {G.~P.}\ \bibnamefont
  {{Holder}}}, \bibinfo {author} {\bibfnamefont {K.~A.}\ \bibnamefont
  {{Aird}}}, \bibinfo {author} {\bibfnamefont {R.}~\bibnamefont {{Armstrong}}},
  \bibinfo {author} {\bibfnamefont {M.~L.~N.}\ \bibnamefont {{Ashby}}},
  \bibinfo {author} {\bibfnamefont {M.~R.}\ \bibnamefont {{Becker}}}, \bibinfo
  {author} {\bibfnamefont {B.~A.}\ \bibnamefont {{Benson}}}, \bibinfo {author}
  {\bibfnamefont {T.}~\bibnamefont {{Biesiadzinski}}}, \bibinfo {author}
  {\bibfnamefont {M.}~\bibnamefont {{Brodwin}}}, \bibinfo {author}
  {\bibfnamefont {M.~T.}\ \bibnamefont {{Busha}}}, \bibinfo {author}
  {\bibfnamefont {J.~E.}\ \bibnamefont {{Carlstrom}}}, \bibinfo {author}
  {\bibfnamefont {C.~L.}\ \bibnamefont {{Chang}}}, \bibinfo {author}
  {\bibfnamefont {H.~M.}\ \bibnamefont {{Cho}}}, \bibinfo {author}
  {\bibfnamefont {T.~M.}\ \bibnamefont {{Crawford}}}, \bibinfo {author}
  {\bibfnamefont {A.~T.}\ \bibnamefont {{Crites}}}, \bibinfo {author}
  {\bibfnamefont {T.}~\bibnamefont {{de Haan}}}, \bibinfo {author}
  {\bibfnamefont {S.}~\bibnamefont {{Desai}}}, \bibinfo {author} {\bibfnamefont
  {M.~A.}\ \bibnamefont {{Dobbs}}}, \bibinfo {author} {\bibfnamefont
  {O.}~\bibnamefont {{Dor{\'e}}}}, \bibinfo {author} {\bibfnamefont
  {J.}~\bibnamefont {{Dudley}}}, \bibinfo {author} {\bibfnamefont {J.~E.}\
  \bibnamefont {{Geach}}}, \bibinfo {author} {\bibfnamefont {E.~M.}\
  \bibnamefont {{George}}}, \bibinfo {author} {\bibfnamefont {M.~D.}\
  \bibnamefont {{Gladders}}}, \bibinfo {author} {\bibfnamefont {A.~H.}\
  \bibnamefont {{Gonzalez}}}, \bibinfo {author} {\bibfnamefont {N.~W.}\
  \bibnamefont {{Halverson}}}, \bibinfo {author} {\bibfnamefont
  {N.}~\bibnamefont {{Harrington}}}, \bibinfo {author} {\bibfnamefont {F.~W.}\
  \bibnamefont {{High}}}, \bibinfo {author} {\bibfnamefont {B.~P.}\
  \bibnamefont {{Holden}}}, \bibinfo {author} {\bibfnamefont {W.~L.}\
  \bibnamefont {{Holzapfel}}}, \bibinfo {author} {\bibfnamefont
  {S.}~\bibnamefont {{Hoover}}}, \bibinfo {author} {\bibfnamefont {J.~D.}\
  \bibnamefont {{Hrubes}}}, \bibinfo {author} {\bibfnamefont {M.}~\bibnamefont
  {{Joy}}}, \bibinfo {author} {\bibfnamefont {R.}~\bibnamefont {{Keisler}}},
  \bibinfo {author} {\bibfnamefont {L.}~\bibnamefont {{Knox}}}, \bibinfo
  {author} {\bibfnamefont {A.~T.}\ \bibnamefont {{Lee}}}, \bibinfo {author}
  {\bibfnamefont {E.~M.}\ \bibnamefont {{Leitch}}}, \bibinfo {author}
  {\bibfnamefont {M.}~\bibnamefont {{Lueker}}}, \bibinfo {author}
  {\bibfnamefont {D.}~\bibnamefont {{Luong-Van}}}, \bibinfo {author}
  {\bibfnamefont {D.~P.}\ \bibnamefont {{Marrone}}}, \bibinfo {author}
  {\bibfnamefont {J.}~\bibnamefont {{Martinez-Manso}}}, \bibinfo {author}
  {\bibfnamefont {J.~J.}\ \bibnamefont {{McMahon}}}, \bibinfo {author}
  {\bibfnamefont {J.}~\bibnamefont {{Mehl}}}, \bibinfo {author} {\bibfnamefont
  {S.~S.}\ \bibnamefont {{Meyer}}}, \bibinfo {author} {\bibfnamefont {J.~J.}\
  \bibnamefont {{Mohr}}}, \bibinfo {author} {\bibfnamefont {T.~E.}\
  \bibnamefont {{Montroy}}}, \bibinfo {author} {\bibfnamefont {T.}~\bibnamefont
  {{Natoli}}}, \bibinfo {author} {\bibfnamefont {S.}~\bibnamefont {{Padin}}},
  \bibinfo {author} {\bibfnamefont {T.}~\bibnamefont {{Plagge}}}, \bibinfo
  {author} {\bibfnamefont {C.}~\bibnamefont {{Pryke}}}, \bibinfo {author}
  {\bibfnamefont {C.~L.}\ \bibnamefont {{Reichardt}}}, \bibinfo {author}
  {\bibfnamefont {A.}~\bibnamefont {{Rest}}}, \bibinfo {author} {\bibfnamefont
  {J.~E.}\ \bibnamefont {{Ruhl}}}, \bibinfo {author} {\bibfnamefont {B.~R.}\
  \bibnamefont {{Saliwanchik}}}, \bibinfo {author} {\bibfnamefont {J.~T.}\
  \bibnamefont {{Sayre}}}, \bibinfo {author} {\bibfnamefont {K.~K.}\
  \bibnamefont {{Schaffer}}}, \bibinfo {author} {\bibfnamefont
  {L.}~\bibnamefont {{Shaw}}}, \bibinfo {author} {\bibfnamefont
  {E.}~\bibnamefont {{Shirokoff}}}, \bibinfo {author} {\bibfnamefont {H.~G.}\
  \bibnamefont {{Spieler}}}, \bibinfo {author} {\bibfnamefont {B.}~\bibnamefont
  {{Stalder}}}, \bibinfo {author} {\bibfnamefont {S.~A.}\ \bibnamefont
  {{Stanford}}}, \bibinfo {author} {\bibfnamefont {Z.}~\bibnamefont
  {{Staniszewski}}}, \bibinfo {author} {\bibfnamefont {A.~A.}\ \bibnamefont
  {{Stark}}}, \bibinfo {author} {\bibfnamefont {D.}~\bibnamefont {{Stern}}},
  \bibinfo {author} {\bibfnamefont {K.}~\bibnamefont {{Story}}}, \bibinfo
  {author} {\bibfnamefont {A.}~\bibnamefont {{Vallinotto}}}, \bibinfo {author}
  {\bibfnamefont {K.}~\bibnamefont {{Vanderlinde}}}, \bibinfo {author}
  {\bibfnamefont {J.~D.}\ \bibnamefont {{Vieira}}}, \bibinfo {author}
  {\bibfnamefont {R.~H.}\ \bibnamefont {{Wechsler}}}, \bibinfo {author}
  {\bibfnamefont {R.}~\bibnamefont {{Williamson}}}, \ and\ \bibinfo {author}
  {\bibfnamefont {O.}~\bibnamefont {{Zahn}}},\ }\href {\doibase
  10.1088/2041-8205/753/1/L9} {\bibfield  {journal} {\bibinfo  {journal}
  {\apjl}\ }\textbf {\bibinfo {volume} {753}},\ \bibinfo {eid} {L9} (\bibinfo
  {year} {2012})},\ \Eprint {http://arxiv.org/abs/1203.4808} {arXiv:1203.4808
  [astro-ph.CO]} \BibitemShut {NoStop}%
\bibitem [{\citenamefont {{Feng}}\ \emph {et~al.}(2012)\citenamefont {{Feng}},
  \citenamefont {{Aslanyan}}, \citenamefont {{Manohar}}, \citenamefont
  {{Keating}}, \citenamefont {{Paar}},\ and\ \citenamefont
  {{Zahn}}}]{2012PhRvD..86f3519F}%
  \BibitemOpen
  \bibfield  {author} {\bibinfo {author} {\bibfnamefont {C.}~\bibnamefont
  {{Feng}}}, \bibinfo {author} {\bibfnamefont {G.}~\bibnamefont {{Aslanyan}}},
  \bibinfo {author} {\bibfnamefont {A.~V.}\ \bibnamefont {{Manohar}}}, \bibinfo
  {author} {\bibfnamefont {B.}~\bibnamefont {{Keating}}}, \bibinfo {author}
  {\bibfnamefont {H.~P.}\ \bibnamefont {{Paar}}}, \ and\ \bibinfo {author}
  {\bibfnamefont {O.}~\bibnamefont {{Zahn}}},\ }\href {\doibase
  10.1103/PhysRevD.86.063519} {\bibfield  {journal} {\bibinfo  {journal}
  {\prd}\ }\textbf {\bibinfo {volume} {86}},\ \bibinfo {eid} {063519} (\bibinfo
  {year} {2012})},\ \Eprint {http://arxiv.org/abs/1207.3326} {arXiv:1207.3326
  [astro-ph.CO]} \BibitemShut {NoStop}%
\bibitem [{\citenamefont {{Sherwin}}\ \emph {et~al.}(2012)\citenamefont
  {{Sherwin}}, \citenamefont {{Das}}, \citenamefont {{Hajian}}, \citenamefont
  {{Addison}}, \citenamefont {{Bond}}, \citenamefont {{Crichton}},
  \citenamefont {{Devlin}}, \citenamefont {{Dunkley}}, \citenamefont
  {{Gralla}}, \citenamefont {{Halpern}}, \citenamefont {{Hill}}, \citenamefont
  {{Hincks}}, \citenamefont {{Hughes}}, \citenamefont {{Huffenberger}},
  \citenamefont {{Hlozek}}, \citenamefont {{Kosowsky}}, \citenamefont
  {{Louis}}, \citenamefont {{Marriage}}, \citenamefont {{Marsden}},
  \citenamefont {{Menanteau}}, \citenamefont {{Moodley}}, \citenamefont
  {{Niemack}}, \citenamefont {{Page}}, \citenamefont {{Reese}}, \citenamefont
  {{Sehgal}}, \citenamefont {{Sievers}}, \citenamefont {{Sif{\'o}n}},
  \citenamefont {{Spergel}}, \citenamefont {{Staggs}}, \citenamefont
  {{Switzer}},\ and\ \citenamefont {{Wollack}}}]{2012PhRvD..86h3006S}%
  \BibitemOpen
  \bibfield  {author} {\bibinfo {author} {\bibfnamefont {B.~D.}\ \bibnamefont
  {{Sherwin}}}, \bibinfo {author} {\bibfnamefont {S.}~\bibnamefont {{Das}}},
  \bibinfo {author} {\bibfnamefont {A.}~\bibnamefont {{Hajian}}}, \bibinfo
  {author} {\bibfnamefont {G.}~\bibnamefont {{Addison}}}, \bibinfo {author}
  {\bibfnamefont {J.~R.}\ \bibnamefont {{Bond}}}, \bibinfo {author}
  {\bibfnamefont {D.}~\bibnamefont {{Crichton}}}, \bibinfo {author}
  {\bibfnamefont {M.~J.}\ \bibnamefont {{Devlin}}}, \bibinfo {author}
  {\bibfnamefont {J.}~\bibnamefont {{Dunkley}}}, \bibinfo {author}
  {\bibfnamefont {M.~B.}\ \bibnamefont {{Gralla}}}, \bibinfo {author}
  {\bibfnamefont {M.}~\bibnamefont {{Halpern}}}, \bibinfo {author}
  {\bibfnamefont {J.~C.}\ \bibnamefont {{Hill}}}, \bibinfo {author}
  {\bibfnamefont {A.~D.}\ \bibnamefont {{Hincks}}}, \bibinfo {author}
  {\bibfnamefont {J.~P.}\ \bibnamefont {{Hughes}}}, \bibinfo {author}
  {\bibfnamefont {K.}~\bibnamefont {{Huffenberger}}}, \bibinfo {author}
  {\bibfnamefont {R.}~\bibnamefont {{Hlozek}}}, \bibinfo {author}
  {\bibfnamefont {A.}~\bibnamefont {{Kosowsky}}}, \bibinfo {author}
  {\bibfnamefont {T.}~\bibnamefont {{Louis}}}, \bibinfo {author} {\bibfnamefont
  {T.~A.}\ \bibnamefont {{Marriage}}}, \bibinfo {author} {\bibfnamefont
  {D.}~\bibnamefont {{Marsden}}}, \bibinfo {author} {\bibfnamefont
  {F.}~\bibnamefont {{Menanteau}}}, \bibinfo {author} {\bibfnamefont
  {K.}~\bibnamefont {{Moodley}}}, \bibinfo {author} {\bibfnamefont {M.~D.}\
  \bibnamefont {{Niemack}}}, \bibinfo {author} {\bibfnamefont {L.~A.}\
  \bibnamefont {{Page}}}, \bibinfo {author} {\bibfnamefont {E.~D.}\
  \bibnamefont {{Reese}}}, \bibinfo {author} {\bibfnamefont {N.}~\bibnamefont
  {{Sehgal}}}, \bibinfo {author} {\bibfnamefont {J.}~\bibnamefont {{Sievers}}},
  \bibinfo {author} {\bibfnamefont {C.}~\bibnamefont {{Sif{\'o}n}}}, \bibinfo
  {author} {\bibfnamefont {D.~N.}\ \bibnamefont {{Spergel}}}, \bibinfo {author}
  {\bibfnamefont {S.~T.}\ \bibnamefont {{Staggs}}}, \bibinfo {author}
  {\bibfnamefont {E.~R.}\ \bibnamefont {{Switzer}}}, \ and\ \bibinfo {author}
  {\bibfnamefont {E.}~\bibnamefont {{Wollack}}},\ }\href {\doibase
  10.1103/PhysRevD.86.083006} {\bibfield  {journal} {\bibinfo  {journal}
  {\prd}\ }\textbf {\bibinfo {volume} {86}},\ \bibinfo {eid} {083006} (\bibinfo
  {year} {2012})},\ \Eprint {http://arxiv.org/abs/1207.4543} {arXiv:1207.4543
  [astro-ph.CO]} \BibitemShut {NoStop}%
\bibitem [{\citenamefont {{Planck Collaboration}}\ \emph
  {et~al.}(2016)\citenamefont {{Planck Collaboration}}, \citenamefont {{Ade}},
  \citenamefont {{Aghanim}}, \citenamefont {{Arnaud}}, \citenamefont
  {{Ashdown}}, \citenamefont {{Aumont}}, \citenamefont {{Baccigalupi}},
  \citenamefont {{Banday}}, \citenamefont {{Barreiro}}, \citenamefont
  {{Bartlett}},\ and\ \citenamefont {et~al.}}]{2016A&A...594A..15P}%
  \BibitemOpen
  \bibfield  {author} {\bibinfo {author} {\bibnamefont {{Planck
  Collaboration}}}, \bibinfo {author} {\bibfnamefont {P.~A.~R.}\ \bibnamefont
  {{Ade}}}, \bibinfo {author} {\bibfnamefont {N.}~\bibnamefont {{Aghanim}}},
  \bibinfo {author} {\bibfnamefont {M.}~\bibnamefont {{Arnaud}}}, \bibinfo
  {author} {\bibfnamefont {M.}~\bibnamefont {{Ashdown}}}, \bibinfo {author}
  {\bibfnamefont {J.}~\bibnamefont {{Aumont}}}, \bibinfo {author}
  {\bibfnamefont {C.}~\bibnamefont {{Baccigalupi}}}, \bibinfo {author}
  {\bibfnamefont {A.~J.}\ \bibnamefont {{Banday}}}, \bibinfo {author}
  {\bibfnamefont {R.~B.}\ \bibnamefont {{Barreiro}}}, \bibinfo {author}
  {\bibfnamefont {J.~G.}\ \bibnamefont {{Bartlett}}}, \ and\ \bibinfo {author}
  {\bibnamefont {et~al.}},\ }\href {\doibase 10.1051/0004-6361/201525941}
  {\bibfield  {journal} {\bibinfo  {journal} {\aap}\ }\textbf {\bibinfo
  {volume} {594}},\ \bibinfo {eid} {A15} (\bibinfo {year} {2016})},\ \Eprint
  {http://arxiv.org/abs/1502.01591} {arXiv:1502.01591} \BibitemShut {NoStop}%
\bibitem [{\citenamefont {Hu}(1999)}]{Hu:1999ek}%
  \BibitemOpen
  \bibfield  {author} {\bibinfo {author} {\bibfnamefont {W.}~\bibnamefont
  {Hu}},\ }\href {\doibase 10.1086/312210} {\bibfield  {journal} {\bibinfo
  {journal} {Astrophys. J.}\ }\textbf {\bibinfo {volume} {522}},\ \bibinfo
  {pages} {L21} (\bibinfo {year} {1999})},\ \Eprint
  {http://arxiv.org/abs/astro-ph/9904153} {arXiv:astro-ph/9904153 [astro-ph]}
  \BibitemShut {NoStop}%
\bibitem [{\citenamefont {{de Putter}}\ \emph {et~al.}(2009)\citenamefont {{de
  Putter}}, \citenamefont {{Zahn}},\ and\ \citenamefont
  {{Linder}}}]{2009PhRvD..79f5033D}%
  \BibitemOpen
  \bibfield  {author} {\bibinfo {author} {\bibfnamefont {R.}~\bibnamefont {{de
  Putter}}}, \bibinfo {author} {\bibfnamefont {O.}~\bibnamefont {{Zahn}}}, \
  and\ \bibinfo {author} {\bibfnamefont {E.~V.}\ \bibnamefont {{Linder}}},\
  }\href {\doibase 10.1103/PhysRevD.79.065033} {\bibfield  {journal} {\bibinfo
  {journal} {\prd}\ }\textbf {\bibinfo {volume} {79}},\ \bibinfo {eid} {065033}
  (\bibinfo {year} {2009})},\ \Eprint {http://arxiv.org/abs/0901.0916}
  {arXiv:0901.0916 [astro-ph.CO]} \BibitemShut {NoStop}%
\bibitem [{\citenamefont {{Das}}\ \emph {et~al.}(2012)\citenamefont {{Das}},
  \citenamefont {{de Putter}}, \citenamefont {{Linder}},\ and\ \citenamefont
  {{Nakajima}}}]{2012JCAP...11..011D}%
  \BibitemOpen
  \bibfield  {author} {\bibinfo {author} {\bibfnamefont {S.}~\bibnamefont
  {{Das}}}, \bibinfo {author} {\bibfnamefont {R.}~\bibnamefont {{de Putter}}},
  \bibinfo {author} {\bibfnamefont {E.~V.}\ \bibnamefont {{Linder}}}, \ and\
  \bibinfo {author} {\bibfnamefont {R.}~\bibnamefont {{Nakajima}}},\ }\href
  {\doibase 10.1088/1475-7516/2012/11/011} {\bibfield  {journal} {\bibinfo
  {journal} {\jcap}\ }\textbf {\bibinfo {volume} {11}},\ \bibinfo {eid} {011}
  (\bibinfo {year} {2012})},\ \Eprint {http://arxiv.org/abs/1102.5090}
  {arXiv:1102.5090 [astro-ph.CO]} \BibitemShut {NoStop}%
\bibitem [{\citenamefont {{Lewis}}\ and\ \citenamefont
  {{Challinor}}(2006)}]{2006PhR...429....1L}%
  \BibitemOpen
  \bibfield  {author} {\bibinfo {author} {\bibfnamefont {A.}~\bibnamefont
  {{Lewis}}}\ and\ \bibinfo {author} {\bibfnamefont {A.}~\bibnamefont
  {{Challinor}}},\ }\href {\doibase 10.1016/j.physrep.2006.03.002} {\bibfield
  {journal} {\bibinfo  {journal} {\physrep}\ }\textbf {\bibinfo {volume}
  {429}},\ \bibinfo {pages} {1} (\bibinfo {year} {2006})},\ \Eprint
  {http://arxiv.org/abs/astro-ph/0601594} {astro-ph/0601594} \BibitemShut
  {NoStop}%
\bibitem [{\citenamefont {{Munshi}}\ \emph {et~al.}(2008)\citenamefont
  {{Munshi}}, \citenamefont {{Valageas}}, \citenamefont {{van Waerbeke}},\ and\
  \citenamefont {{Heavens}}}]{2008PhR...462...67M}%
  \BibitemOpen
  \bibfield  {author} {\bibinfo {author} {\bibfnamefont {D.}~\bibnamefont
  {{Munshi}}}, \bibinfo {author} {\bibfnamefont {P.}~\bibnamefont
  {{Valageas}}}, \bibinfo {author} {\bibfnamefont {L.}~\bibnamefont {{van
  Waerbeke}}}, \ and\ \bibinfo {author} {\bibfnamefont {A.}~\bibnamefont
  {{Heavens}}},\ }\href {\doibase 10.1016/j.physrep.2008.02.003} {\bibfield
  {journal} {\bibinfo  {journal} {\physrep}\ }\textbf {\bibinfo {volume}
  {462}},\ \bibinfo {pages} {67} (\bibinfo {year} {2008})},\ \Eprint
  {http://arxiv.org/abs/astro-ph/0612667} {astro-ph/0612667} \BibitemShut
  {NoStop}%
\bibitem [{\citenamefont {Niemack}\ \emph {et~al.}(2010)\citenamefont
  {Niemack}, \citenamefont {Ade}, \citenamefont {Aguirre}, \citenamefont
  {Barrientos}, \citenamefont {Beall}, \citenamefont {Bond}, \citenamefont
  {Britton}, \citenamefont {Cho}, \citenamefont {Das}, \citenamefont {Devlin},
  \citenamefont {Dicker}, \citenamefont {Dunkley}, \citenamefont {Dünner},
  \citenamefont {Fowler}, \citenamefont {Hajian}, \citenamefont {Halpern},
  \citenamefont {Hasselfield}, \citenamefont {Hilton}, \citenamefont {Hilton},
  \citenamefont {Hubmayr}, \citenamefont {Hughes}, \citenamefont {Infante},
  \citenamefont {Irwin}, \citenamefont {Jarosik}, \citenamefont {Klein},
  \citenamefont {Kosowsky}, \citenamefont {Marriage}, \citenamefont {McMahon},
  \citenamefont {Menanteau}, \citenamefont {Moodley}, \citenamefont {Nibarger},
  \citenamefont {Nolta}, \citenamefont {Page}, \citenamefont {Partridge},
  \citenamefont {Reese}, \citenamefont {Sievers}, \citenamefont {Spergel},
  \citenamefont {Staggs}, \citenamefont {Thornton}, \citenamefont {Tucker},
  \citenamefont {Wollack},\ and\ \citenamefont {Yoon}}]{doi:10.1117/12.857464}%
  \BibitemOpen
  \bibfield  {author} {\bibinfo {author} {\bibfnamefont {M.~D.}\ \bibnamefont
  {Niemack}}, \bibinfo {author} {\bibfnamefont {P.~A.~R.}\ \bibnamefont {Ade}},
  \bibinfo {author} {\bibfnamefont {J.}~\bibnamefont {Aguirre}}, \bibinfo
  {author} {\bibfnamefont {F.}~\bibnamefont {Barrientos}}, \bibinfo {author}
  {\bibfnamefont {J.~A.}\ \bibnamefont {Beall}}, \bibinfo {author}
  {\bibfnamefont {J.~R.}\ \bibnamefont {Bond}}, \bibinfo {author}
  {\bibfnamefont {J.}~\bibnamefont {Britton}}, \bibinfo {author} {\bibfnamefont
  {H.~M.}\ \bibnamefont {Cho}}, \bibinfo {author} {\bibfnamefont
  {S.}~\bibnamefont {Das}}, \bibinfo {author} {\bibfnamefont {M.~J.}\
  \bibnamefont {Devlin}}, \bibinfo {author} {\bibfnamefont {S.}~\bibnamefont
  {Dicker}}, \bibinfo {author} {\bibfnamefont {J.}~\bibnamefont {Dunkley}},
  \bibinfo {author} {\bibfnamefont {R.}~\bibnamefont {Dünner}}, \bibinfo
  {author} {\bibfnamefont {J.~W.}\ \bibnamefont {Fowler}}, \bibinfo {author}
  {\bibfnamefont {A.}~\bibnamefont {Hajian}}, \bibinfo {author} {\bibfnamefont
  {M.}~\bibnamefont {Halpern}}, \bibinfo {author} {\bibfnamefont
  {M.}~\bibnamefont {Hasselfield}}, \bibinfo {author} {\bibfnamefont {G.~C.}\
  \bibnamefont {Hilton}}, \bibinfo {author} {\bibfnamefont {M.}~\bibnamefont
  {Hilton}}, \bibinfo {author} {\bibfnamefont {J.}~\bibnamefont {Hubmayr}},
  \bibinfo {author} {\bibfnamefont {J.~P.}\ \bibnamefont {Hughes}}, \bibinfo
  {author} {\bibfnamefont {L.}~\bibnamefont {Infante}}, \bibinfo {author}
  {\bibfnamefont {K.~D.}\ \bibnamefont {Irwin}}, \bibinfo {author}
  {\bibfnamefont {N.}~\bibnamefont {Jarosik}}, \bibinfo {author} {\bibfnamefont
  {J.}~\bibnamefont {Klein}}, \bibinfo {author} {\bibfnamefont
  {A.}~\bibnamefont {Kosowsky}}, \bibinfo {author} {\bibfnamefont {T.~A.}\
  \bibnamefont {Marriage}}, \bibinfo {author} {\bibfnamefont {J.}~\bibnamefont
  {McMahon}}, \bibinfo {author} {\bibfnamefont {F.}~\bibnamefont {Menanteau}},
  \bibinfo {author} {\bibfnamefont {K.}~\bibnamefont {Moodley}}, \bibinfo
  {author} {\bibfnamefont {J.~P.}\ \bibnamefont {Nibarger}}, \bibinfo {author}
  {\bibfnamefont {M.~R.}\ \bibnamefont {Nolta}}, \bibinfo {author}
  {\bibfnamefont {L.~A.}\ \bibnamefont {Page}}, \bibinfo {author}
  {\bibfnamefont {B.}~\bibnamefont {Partridge}}, \bibinfo {author}
  {\bibfnamefont {E.~D.}\ \bibnamefont {Reese}}, \bibinfo {author}
  {\bibfnamefont {J.}~\bibnamefont {Sievers}}, \bibinfo {author} {\bibfnamefont
  {D.~N.}\ \bibnamefont {Spergel}}, \bibinfo {author} {\bibfnamefont {S.~T.}\
  \bibnamefont {Staggs}}, \bibinfo {author} {\bibfnamefont {R.}~\bibnamefont
  {Thornton}}, \bibinfo {author} {\bibfnamefont {C.}~\bibnamefont {Tucker}},
  \bibinfo {author} {\bibfnamefont {E.}~\bibnamefont {Wollack}}, \ and\
  \bibinfo {author} {\bibfnamefont {K.~W.}\ \bibnamefont {Yoon}},\ }\href
  {\doibase 10.1117/12.857464} {\bibfield  {journal} {\bibinfo  {journal}
  {Proc. SPIE}\ }\textbf {\bibinfo {volume} {7741}},\ \bibinfo {pages} {77411S}
  (\bibinfo {year} {2010})}\BibitemShut {NoStop}%
\bibitem [{\citenamefont {{Austermann}}\ \emph {et~al.}(2012)\citenamefont
  {{Austermann}}, \citenamefont {{Aird}}, \citenamefont {{Beall}},
  \citenamefont {{Becker}}, \citenamefont {{Bender}}, \citenamefont {{Benson}},
  \citenamefont {{Bleem}}, \citenamefont {{Britton}}, \citenamefont
  {{Carlstrom}}, \citenamefont {{Chang}}, \citenamefont {{Chiang}},
  \citenamefont {{Cho}}, \citenamefont {{Crawford}}, \citenamefont {{Crites}},
  \citenamefont {{Datesman}}, \citenamefont {{de Haan}}, \citenamefont
  {{Dobbs}}, \citenamefont {{George}}, \citenamefont {{Halverson}},
  \citenamefont {{Harrington}}, \citenamefont {{Henning}}, \citenamefont
  {{Hilton}}, \citenamefont {{Holder}}, \citenamefont {{Holzapfel}},
  \citenamefont {{Hoover}}, \citenamefont {{Huang}}, \citenamefont {{Hubmayr}},
  \citenamefont {{Irwin}}, \citenamefont {{Keisler}}, \citenamefont
  {{Kennedy}}, \citenamefont {{Knox}}, \citenamefont {{Lee}}, \citenamefont
  {{Leitch}}, \citenamefont {{Li}}, \citenamefont {{Lueker}}, \citenamefont
  {{Marrone}}, \citenamefont {{McMahon}}, \citenamefont {{Mehl}}, \citenamefont
  {{Meyer}}, \citenamefont {{Montroy}}, \citenamefont {{Natoli}}, \citenamefont
  {{Nibarger}}, \citenamefont {{Niemack}}, \citenamefont {{Novosad}},
  \citenamefont {{Padin}}, \citenamefont {{Pryke}}, \citenamefont
  {{Reichardt}}, \citenamefont {{Ruhl}}, \citenamefont {{Saliwanchik}},
  \citenamefont {{Sayre}}, \citenamefont {{Schaffer}}, \citenamefont
  {{Shirokoff}}, \citenamefont {{Stark}}, \citenamefont {{Story}},
  \citenamefont {{Vanderlinde}}, \citenamefont {{Vieira}}, \citenamefont
  {{Wang}}, \citenamefont {{Williamson}}, \citenamefont {{Yefremenko}},
  \citenamefont {{Yoon}},\ and\ \citenamefont {{Zahn}}}]{2012SPIE.8452E..1EA}%
  \BibitemOpen
  \bibfield  {author} {\bibinfo {author} {\bibfnamefont {J.~E.}\ \bibnamefont
  {{Austermann}}}, \bibinfo {author} {\bibfnamefont {K.~A.}\ \bibnamefont
  {{Aird}}}, \bibinfo {author} {\bibfnamefont {J.~A.}\ \bibnamefont {{Beall}}},
  \bibinfo {author} {\bibfnamefont {D.}~\bibnamefont {{Becker}}}, \bibinfo
  {author} {\bibfnamefont {A.}~\bibnamefont {{Bender}}}, \bibinfo {author}
  {\bibfnamefont {B.~A.}\ \bibnamefont {{Benson}}}, \bibinfo {author}
  {\bibfnamefont {L.~E.}\ \bibnamefont {{Bleem}}}, \bibinfo {author}
  {\bibfnamefont {J.}~\bibnamefont {{Britton}}}, \bibinfo {author}
  {\bibfnamefont {J.~E.}\ \bibnamefont {{Carlstrom}}}, \bibinfo {author}
  {\bibfnamefont {C.~L.}\ \bibnamefont {{Chang}}}, \bibinfo {author}
  {\bibfnamefont {H.~C.}\ \bibnamefont {{Chiang}}}, \bibinfo {author}
  {\bibfnamefont {H.-M.}\ \bibnamefont {{Cho}}}, \bibinfo {author}
  {\bibfnamefont {T.~M.}\ \bibnamefont {{Crawford}}}, \bibinfo {author}
  {\bibfnamefont {A.~T.}\ \bibnamefont {{Crites}}}, \bibinfo {author}
  {\bibfnamefont {A.}~\bibnamefont {{Datesman}}}, \bibinfo {author}
  {\bibfnamefont {T.}~\bibnamefont {{de Haan}}}, \bibinfo {author}
  {\bibfnamefont {M.~A.}\ \bibnamefont {{Dobbs}}}, \bibinfo {author}
  {\bibfnamefont {E.~M.}\ \bibnamefont {{George}}}, \bibinfo {author}
  {\bibfnamefont {N.~W.}\ \bibnamefont {{Halverson}}}, \bibinfo {author}
  {\bibfnamefont {N.}~\bibnamefont {{Harrington}}}, \bibinfo {author}
  {\bibfnamefont {J.~W.}\ \bibnamefont {{Henning}}}, \bibinfo {author}
  {\bibfnamefont {G.~C.}\ \bibnamefont {{Hilton}}}, \bibinfo {author}
  {\bibfnamefont {G.~P.}\ \bibnamefont {{Holder}}}, \bibinfo {author}
  {\bibfnamefont {W.~L.}\ \bibnamefont {{Holzapfel}}}, \bibinfo {author}
  {\bibfnamefont {S.}~\bibnamefont {{Hoover}}}, \bibinfo {author}
  {\bibfnamefont {N.}~\bibnamefont {{Huang}}}, \bibinfo {author} {\bibfnamefont
  {J.}~\bibnamefont {{Hubmayr}}}, \bibinfo {author} {\bibfnamefont {K.~D.}\
  \bibnamefont {{Irwin}}}, \bibinfo {author} {\bibfnamefont {R.}~\bibnamefont
  {{Keisler}}}, \bibinfo {author} {\bibfnamefont {J.}~\bibnamefont
  {{Kennedy}}}, \bibinfo {author} {\bibfnamefont {L.}~\bibnamefont {{Knox}}},
  \bibinfo {author} {\bibfnamefont {A.~T.}\ \bibnamefont {{Lee}}}, \bibinfo
  {author} {\bibfnamefont {E.}~\bibnamefont {{Leitch}}}, \bibinfo {author}
  {\bibfnamefont {D.}~\bibnamefont {{Li}}}, \bibinfo {author} {\bibfnamefont
  {M.}~\bibnamefont {{Lueker}}}, \bibinfo {author} {\bibfnamefont {D.~P.}\
  \bibnamefont {{Marrone}}}, \bibinfo {author} {\bibfnamefont {J.~J.}\
  \bibnamefont {{McMahon}}}, \bibinfo {author} {\bibfnamefont {J.}~\bibnamefont
  {{Mehl}}}, \bibinfo {author} {\bibfnamefont {S.~S.}\ \bibnamefont {{Meyer}}},
  \bibinfo {author} {\bibfnamefont {T.~E.}\ \bibnamefont {{Montroy}}}, \bibinfo
  {author} {\bibfnamefont {T.}~\bibnamefont {{Natoli}}}, \bibinfo {author}
  {\bibfnamefont {J.~P.}\ \bibnamefont {{Nibarger}}}, \bibinfo {author}
  {\bibfnamefont {M.~D.}\ \bibnamefont {{Niemack}}}, \bibinfo {author}
  {\bibfnamefont {V.}~\bibnamefont {{Novosad}}}, \bibinfo {author}
  {\bibfnamefont {S.}~\bibnamefont {{Padin}}}, \bibinfo {author} {\bibfnamefont
  {C.}~\bibnamefont {{Pryke}}}, \bibinfo {author} {\bibfnamefont {C.~L.}\
  \bibnamefont {{Reichardt}}}, \bibinfo {author} {\bibfnamefont {J.~E.}\
  \bibnamefont {{Ruhl}}}, \bibinfo {author} {\bibfnamefont {B.~R.}\
  \bibnamefont {{Saliwanchik}}}, \bibinfo {author} {\bibfnamefont {J.~T.}\
  \bibnamefont {{Sayre}}}, \bibinfo {author} {\bibfnamefont {K.~K.}\
  \bibnamefont {{Schaffer}}}, \bibinfo {author} {\bibfnamefont
  {E.}~\bibnamefont {{Shirokoff}}}, \bibinfo {author} {\bibfnamefont {A.~A.}\
  \bibnamefont {{Stark}}}, \bibinfo {author} {\bibfnamefont {K.}~\bibnamefont
  {{Story}}}, \bibinfo {author} {\bibfnamefont {K.}~\bibnamefont
  {{Vanderlinde}}}, \bibinfo {author} {\bibfnamefont {J.~D.}\ \bibnamefont
  {{Vieira}}}, \bibinfo {author} {\bibfnamefont {G.}~\bibnamefont {{Wang}}},
  \bibinfo {author} {\bibfnamefont {R.}~\bibnamefont {{Williamson}}}, \bibinfo
  {author} {\bibfnamefont {V.}~\bibnamefont {{Yefremenko}}}, \bibinfo {author}
  {\bibfnamefont {K.~W.}\ \bibnamefont {{Yoon}}}, \ and\ \bibinfo {author}
  {\bibfnamefont {O.}~\bibnamefont {{Zahn}}},\ }in\ \href {\doibase
  10.1117/12.927286} {\emph {\bibinfo {booktitle} {Millimeter, Submillimeter,
  and Far-Infrared Detectors and Instrumentation for Astronomy VI}}},\ \bibinfo
  {series} {\procspie}, Vol.\ \bibinfo {volume} {8452}\ (\bibinfo {year}
  {2012})\ p.\ \bibinfo {pages} {84521E},\ \Eprint
  {http://arxiv.org/abs/1210.4970} {arXiv:1210.4970 [astro-ph.IM]} \BibitemShut
  {NoStop}%
\bibitem [{\citenamefont {Arnold}\ \emph {et~al.}(2010)\citenamefont {Arnold},
  \citenamefont {Ade}, \citenamefont {Anthony}, \citenamefont {Aubin},
  \citenamefont {Boettger}, \citenamefont {Borrill}, \citenamefont {Cantalupo},
  \citenamefont {Dobbs}, \citenamefont {Errard}, \citenamefont {Flanigan},
  \citenamefont {Ghribi}, \citenamefont {Halverson}, \citenamefont {Hazumi},
  \citenamefont {Holzapfel}, \citenamefont {Howard}, \citenamefont {Hyland},
  \citenamefont {Jaffe}, \citenamefont {Keating}, \citenamefont {Kisner},
  \citenamefont {Kermish}, \citenamefont {Lee}, \citenamefont {Linder},
  \citenamefont {Lungu}, \citenamefont {Matsumura}, \citenamefont {Miller},
  \citenamefont {Meng}, \citenamefont {Myers}, \citenamefont {Nishino},
  \citenamefont {O'Brient}, \citenamefont {O'Dea}, \citenamefont {Paar},
  \citenamefont {Reichardt}, \citenamefont {Schanning}, \citenamefont
  {Shimizu}, \citenamefont {Shimmin}, \citenamefont {Shimon}, \citenamefont
  {Spieler}, \citenamefont {Steinbach}, \citenamefont {Stompor}, \citenamefont
  {Suzuki}, \citenamefont {Tomaru}, \citenamefont {Tran}, \citenamefont
  {Tucker}, \citenamefont {Quealy}, \citenamefont {Richards},\ and\
  \citenamefont {Zahn}}]{doi:10.1117/12.858314}%
  \BibitemOpen
  \bibfield  {author} {\bibinfo {author} {\bibfnamefont {K.}~\bibnamefont
  {Arnold}}, \bibinfo {author} {\bibfnamefont {P.~A.~R.}\ \bibnamefont {Ade}},
  \bibinfo {author} {\bibfnamefont {A.~E.}\ \bibnamefont {Anthony}}, \bibinfo
  {author} {\bibfnamefont {F.}~\bibnamefont {Aubin}}, \bibinfo {author}
  {\bibfnamefont {D.}~\bibnamefont {Boettger}}, \bibinfo {author}
  {\bibfnamefont {J.}~\bibnamefont {Borrill}}, \bibinfo {author} {\bibfnamefont
  {C.}~\bibnamefont {Cantalupo}}, \bibinfo {author} {\bibfnamefont {M.~A.}\
  \bibnamefont {Dobbs}}, \bibinfo {author} {\bibfnamefont {J.}~\bibnamefont
  {Errard}}, \bibinfo {author} {\bibfnamefont {D.}~\bibnamefont {Flanigan}},
  \bibinfo {author} {\bibfnamefont {A.}~\bibnamefont {Ghribi}}, \bibinfo
  {author} {\bibfnamefont {N.}~\bibnamefont {Halverson}}, \bibinfo {author}
  {\bibfnamefont {M.}~\bibnamefont {Hazumi}}, \bibinfo {author} {\bibfnamefont
  {W.~L.}\ \bibnamefont {Holzapfel}}, \bibinfo {author} {\bibfnamefont
  {J.}~\bibnamefont {Howard}}, \bibinfo {author} {\bibfnamefont
  {P.}~\bibnamefont {Hyland}}, \bibinfo {author} {\bibfnamefont
  {A.}~\bibnamefont {Jaffe}}, \bibinfo {author} {\bibfnamefont
  {B.}~\bibnamefont {Keating}}, \bibinfo {author} {\bibfnamefont
  {T.}~\bibnamefont {Kisner}}, \bibinfo {author} {\bibfnamefont
  {Z.}~\bibnamefont {Kermish}}, \bibinfo {author} {\bibfnamefont {A.~T.}\
  \bibnamefont {Lee}}, \bibinfo {author} {\bibfnamefont {E.}~\bibnamefont
  {Linder}}, \bibinfo {author} {\bibfnamefont {M.}~\bibnamefont {Lungu}},
  \bibinfo {author} {\bibfnamefont {T.}~\bibnamefont {Matsumura}}, \bibinfo
  {author} {\bibfnamefont {N.}~\bibnamefont {Miller}}, \bibinfo {author}
  {\bibfnamefont {X.}~\bibnamefont {Meng}}, \bibinfo {author} {\bibfnamefont
  {M.}~\bibnamefont {Myers}}, \bibinfo {author} {\bibfnamefont
  {H.}~\bibnamefont {Nishino}}, \bibinfo {author} {\bibfnamefont
  {R.}~\bibnamefont {O'Brient}}, \bibinfo {author} {\bibfnamefont
  {D.}~\bibnamefont {O'Dea}}, \bibinfo {author} {\bibfnamefont
  {H.}~\bibnamefont {Paar}}, \bibinfo {author} {\bibfnamefont {C.}~\bibnamefont
  {Reichardt}}, \bibinfo {author} {\bibfnamefont {I.}~\bibnamefont
  {Schanning}}, \bibinfo {author} {\bibfnamefont {A.}~\bibnamefont {Shimizu}},
  \bibinfo {author} {\bibfnamefont {C.}~\bibnamefont {Shimmin}}, \bibinfo
  {author} {\bibfnamefont {M.}~\bibnamefont {Shimon}}, \bibinfo {author}
  {\bibfnamefont {H.}~\bibnamefont {Spieler}}, \bibinfo {author} {\bibfnamefont
  {B.}~\bibnamefont {Steinbach}}, \bibinfo {author} {\bibfnamefont
  {R.}~\bibnamefont {Stompor}}, \bibinfo {author} {\bibfnamefont
  {A.}~\bibnamefont {Suzuki}}, \bibinfo {author} {\bibfnamefont
  {T.}~\bibnamefont {Tomaru}}, \bibinfo {author} {\bibfnamefont {H.~T.}\
  \bibnamefont {Tran}}, \bibinfo {author} {\bibfnamefont {C.}~\bibnamefont
  {Tucker}}, \bibinfo {author} {\bibfnamefont {E.}~\bibnamefont {Quealy}},
  \bibinfo {author} {\bibfnamefont {P.~L.}\ \bibnamefont {Richards}}, \ and\
  \bibinfo {author} {\bibfnamefont {O.}~\bibnamefont {Zahn}},\ }\href {\doibase
  10.1117/12.858314} {\bibfield  {journal} {\bibinfo  {journal} {Proc. SPIE}\
  }\textbf {\bibinfo {volume} {7741}},\ \bibinfo {pages} {77411E} (\bibinfo
  {year} {2010})}\BibitemShut {NoStop}%
\bibitem [{\citenamefont {{Silk}}(1968)}]{1968ApJ...151..459S}%
  \BibitemOpen
  \bibfield  {author} {\bibinfo {author} {\bibfnamefont {J.}~\bibnamefont
  {{Silk}}},\ }\href {\doibase 10.1086/149449} {\bibfield  {journal} {\bibinfo
  {journal} {\apj}\ }\textbf {\bibinfo {volume} {151}},\ \bibinfo {pages} {459}
  (\bibinfo {year} {1968})}\BibitemShut {NoStop}%
\bibitem [{\citenamefont {{Peebles}}\ and\ \citenamefont
  {{Yu}}(1970)}]{1970ApJ...162..815P}%
  \BibitemOpen
  \bibfield  {author} {\bibinfo {author} {\bibfnamefont {P.~J.~E.}\
  \bibnamefont {{Peebles}}}\ and\ \bibinfo {author} {\bibfnamefont {J.~T.}\
  \bibnamefont {{Yu}}},\ }\href {\doibase 10.1086/150713} {\bibfield  {journal}
  {\bibinfo  {journal} {\apj}\ }\textbf {\bibinfo {volume} {162}},\ \bibinfo
  {pages} {815} (\bibinfo {year} {1970})}\BibitemShut {NoStop}%
\bibitem [{\citenamefont {{Lewis}}\ and\ \citenamefont
  {{Challinor}}(2011)}]{2011ascl.soft02026L}%
  \BibitemOpen
  \bibfield  {author} {\bibinfo {author} {\bibfnamefont {A.}~\bibnamefont
  {{Lewis}}}\ and\ \bibinfo {author} {\bibfnamefont {A.}~\bibnamefont
  {{Challinor}}},\ }\href@noop {} {\enquote {\bibinfo {title} {{CAMB: Code for
  Anisotropies in the Microwave Background}},}\ }\bibinfo {howpublished}
  {Astrophysics Source Code Library} (\bibinfo {year} {2011}),\ \Eprint
  {http://arxiv.org/abs/1102.026} {ascl:1102.026} \BibitemShut {NoStop}%
\bibitem [{\citenamefont {{Blas}}\ \emph {et~al.}(2011)\citenamefont {{Blas}},
  \citenamefont {{Lesgourgues}},\ and\ \citenamefont
  {{Tram}}}]{2011JCAP...07..034B}%
  \BibitemOpen
  \bibfield  {author} {\bibinfo {author} {\bibfnamefont {D.}~\bibnamefont
  {{Blas}}}, \bibinfo {author} {\bibfnamefont {J.}~\bibnamefont
  {{Lesgourgues}}}, \ and\ \bibinfo {author} {\bibfnamefont {T.}~\bibnamefont
  {{Tram}}},\ }\href {\doibase 10.1088/1475-7516/2011/07/034} {\bibfield
  {journal} {\bibinfo  {journal} {\jcap}\ }\textbf {\bibinfo {volume} {7}},\
  \bibinfo {eid} {034} (\bibinfo {year} {2011})},\ \Eprint
  {http://arxiv.org/abs/1104.2933} {arXiv:1104.2933} \BibitemShut {NoStop}%
\bibitem [{\citenamefont {{Bartelmann}}\ and\ \citenamefont
  {{Schneider}}(2001)}]{2001PhR...340..291B}%
  \BibitemOpen
  \bibfield  {author} {\bibinfo {author} {\bibfnamefont {M.}~\bibnamefont
  {{Bartelmann}}}\ and\ \bibinfo {author} {\bibfnamefont {P.}~\bibnamefont
  {{Schneider}}},\ }\href {\doibase 10.1016/S0370-1573(00)00082-X} {\bibfield
  {journal} {\bibinfo  {journal} {\physrep}\ }\textbf {\bibinfo {volume}
  {340}},\ \bibinfo {pages} {291} (\bibinfo {year} {2001})},\ \Eprint
  {http://arxiv.org/abs/astro-ph/9912508} {astro-ph/9912508} \BibitemShut
  {NoStop}%
\bibitem [{\citenamefont {Hu}(1995)}]{Hu:1995em}%
  \BibitemOpen
  \bibfield  {author} {\bibinfo {author} {\bibfnamefont {W.~T.}\ \bibnamefont
  {Hu}},\ }\emph {\bibinfo {title} {{Wandering in the Background: A CMB
  Explorer}}},\ \href {http://alice.cern.ch/format/showfull?sysnb=0207836}
  {Ph.D. thesis},\ \bibinfo  {school} {UC, Berkeley} (\bibinfo {year} {1995}),\
  \Eprint {http://arxiv.org/abs/astro-ph/9508126} {arXiv:astro-ph/9508126
  [astro-ph]} \BibitemShut {NoStop}%
\bibitem [{\citenamefont {{Limber}}(1954)}]{1954ApJ...119..655L}%
  \BibitemOpen
  \bibfield  {author} {\bibinfo {author} {\bibfnamefont {D.~N.}\ \bibnamefont
  {{Limber}}},\ }\href {\doibase 10.1086/145870} {\bibfield  {journal}
  {\bibinfo  {journal} {\apj}\ }\textbf {\bibinfo {volume} {119}},\ \bibinfo
  {pages} {655} (\bibinfo {year} {1954})}\BibitemShut {NoStop}%
\bibitem [{\citenamefont {{Kaiser}}(1992)}]{1992ApJ...388..272K}%
  \BibitemOpen
  \bibfield  {author} {\bibinfo {author} {\bibfnamefont {N.}~\bibnamefont
  {{Kaiser}}},\ }\href {\doibase 10.1086/171151} {\bibfield  {journal}
  {\bibinfo  {journal} {\apj}\ }\textbf {\bibinfo {volume} {388}},\ \bibinfo
  {pages} {272} (\bibinfo {year} {1992})}\BibitemShut {NoStop}%
\bibitem [{\citenamefont {Challinor}\ and\ \citenamefont
  {Lewis}(2005)}]{Challinor:2005jy}%
  \BibitemOpen
  \bibfield  {author} {\bibinfo {author} {\bibfnamefont {A.}~\bibnamefont
  {Challinor}}\ and\ \bibinfo {author} {\bibfnamefont {A.}~\bibnamefont
  {Lewis}},\ }\href {\doibase 10.1103/PhysRevD.71.103010} {\bibfield  {journal}
  {\bibinfo  {journal} {Phys. Rev.}\ }\textbf {\bibinfo {volume} {D71}},\
  \bibinfo {pages} {103010} (\bibinfo {year} {2005})},\ \Eprint
  {http://arxiv.org/abs/astro-ph/0502425} {arXiv:astro-ph/0502425 [astro-ph]}
  \BibitemShut {NoStop}%
\end{thebibliography}%

\end{document}